\documentclass[onecolumn,journal]{IEEEtran}

\IEEEoverridecommandlockouts                              
\usepackage{graphicx}
\usepackage{subfig}
\usepackage{amsmath}
\usepackage{amssymb}
\usepackage[font=footnotesize]{caption} 
\usepackage{subfig} 
\usepackage{cite} 
\usepackage{color}
\usepackage{algorithm} 
\usepackage{algpseudocode} 
\usepackage{enumerate}
\usepackage{hyperref}

\usepackage{setspace}
\usepackage{tikz}
\usetikzlibrary{calc} 

\newtheorem{my_theorem}{Theorem}
\newtheorem{my_lemma}{Lemma}
\newtheorem{my_assumption}{Assumption}
\newtheorem{my_remark}{Remark}

\graphicspath{{figures/}}

\newcommand{\tr}{\mathrm{tr\,}}
\newcommand{\Null}{\mathrm{Null\,}}
\newcommand{\one}{\mathbf{1}}

\newcommand{\sgn}{\mathrm{sgn}}

\begin{document}

\title{\LARGE \bf
Distributed Control of Angle-constrained Circular Formations \\using Bearing-only Measurements
}

\author{Shiyu Zhao,
        Feng Lin,
        Kemao Peng,
        Ben M. Chen
        and Tong H. Lee 
\thanks{S. Zhao, B. M. Chen and T. H. Lee are with the Department of Electrical and Computer Engineering, National University of Singapore, Singapore 117576, Singapore
    {\tt\small \{shiyuzhao, bmchen, eleleeth\}@nus.edu.sg}}
\thanks{F. Lin and K. Peng are with the Temasek Laboratories, National University of Singapore, Singapore 117456, Singapore
    {\tt\small \{linfeng, kmpeng\}@nus.edu.sg}}
}

\maketitle

\begin{abstract}
This paper studies distributed formation control of multiple agents in the plane using bearing-only measurements.
It is assumed that each agent only measures the local bearings of their neighbor agents.
The target formation considered in this paper is a circular formation, where each agent has exactly two neighbors.
In the target formation, the angle subtended at each agent by their two neighbors is specified.
We propose a distributed control law that stabilizes angle-constrained target formations merely using local bearing measurements.
The stability of the target formation is analyzed based on Lyapunov approaches.
We present a unified proof to show that our control law not only can ensure local exponential stability but also can give local finite-time stability.
The exponential or finite-time stability can be easily switched by tuning a parameter in the control law.
\end{abstract}


\IEEEpeerreviewmaketitle

\section{Introduction}

Distributed formation control of multiple agents has been investigated extensively in various settings.
We here would like to highlight two aspects that are important to characterize a formation control problem.

The first aspect is what kind of information each agent can obtain from their neighbors.
In order to realize distributed position control of multiple agents, it is commonly assumed that each agent can obtain the (global or relative) positions of their neighbor agents through wireless communication.
It is interesting to note that the position information consists of two kinds of partial information: distance and bearing.
Formation control merely using partial information has become an active research area in recent years.
The work in \cite{cao2011,cao2012} addresses formation coordination of mobile agents when each agent can only measure the distances to their neighbors.
Formation control using bearing-only measurements has been studied in \cite{bishopSCL,bishopconf2010,bishopconf2011relax,bishopconf2011quad,ErenIJC,nimaconf,nimaTR}.

The second aspect is how the target formation is defined.
Conventionally target formations are defined by specifying global or relative positions of agents.
It is noticed that angles and inter-agent distances can also be used to define a target formation.
The term angle as used here refers to the angle subtended at one agent by its two neighbors.
In recent years, control of distance-constrained formations
has received much attention \cite{reza2002formation,Francis2009IJC,CorteAutomatica2009,Francis2010TAC,yu2010SIAM,Anderson2011TAC,Huang2012}.
Control of angle-constrained formations has also attract some interest very recently
\cite{bishopconf2011rigid,bishopSCL,bishopconf2010,bishopconf2011relax,bishopconf2011quad,ErenIJC}.
For distance-constrained target formations, any rigid body transformation (i.e., rotation and translation) over the entire formation will not change the inter-agent distances.
If the target formation is defined by angles, in addition to rigid body transformation, the target formation will also be invariant to scaling.
Furthermore, without parallel rigidity constraints \cite{bishopconf2011rigid,eren2003,ErenIJC}, the angle-constrained target formation will not be affected by any edge parallel motion either.

We now characterize the formation control problem studied in this paper from the above two aspects.
Our work considers formation control using bearing-only measurements.
That is motivated by an important type of sensor: camera, which inherently is a bearing-only sensor and has been widely applied in many control-related tasks.
Vision-based formation control \cite{nimaconf,nimaTR,TRA2002Vision} is a potential application of our research work.
In this paper, it is assumed that each agent can only measure the bearings of their neighbors in their local coordinate frame.
It should be noted that the bearing measurements are not used to estimate any agent's position. The control is implemented directly based on bearing measurements. 
As a consequence, we can expect that the control is only able to handle angle-constrained target formations.
Specifically our work considers angle-constrained circular target formations, where each agent has exactly two neighbors.
In the target formation, the angle subtended at each agent by their two neighbors is specified.
At last, our work makes no parallel rigidity assumptions \cite{eren2003,bishopconf2011rigid,ErenIJC} of the target formation.

We propose a distributed nonlinear control law that stabilizes angle-constrained target formations merely using local bearing measurements.
The proposed control law is inspired by the work in \cite{bishopSCL,bishopconf2010,bishopconf2011quad} which addressed distributed control of triangular and quadrilateral formations.
An attractive feature of the control law in \cite{bishopSCL,bishopconf2010} is that the global stability can be proved based on the Poincare-Bendixson theorem.
In contrast to previous work, we consider circular target formations with an arbitrary number of agents and the stability by the proposed control law will be analyzed based on Lyapunov approaches.
Our work also involves finite-time formation control which has received some attention recently \cite{wanglong2009finiteformation,caoyongcan2010formation}.
Besides fast convergence, finite-time control can also bring benefits such as disturbance rejection and robustness against uncertainties \cite{bernstein2000}.
We will prove that the proposed control law not only can ensure local exponential stability but also can give local finite-time stability.
The exponential or finite-time stability can be switched by simply tuning a parameter in the control law.

This paper is organized as follows.
Section \ref{section_problem_formulation} formulates the formation control problem and presents our control law.
We prove some useful lemmas in Section \ref{section_useful_lemma}.
The exponential and finite-time stability by the proposed control law is proved in Section \ref{section_stability}.
Formation behaviors including collision avoidance between agents are analyzed in Section \ref{section_formationbehavior}.
Numerical simulations are presented in Section \ref{section_simulation}.
A summary is given in Section \ref{section_conclusion}.

\section{Problem Formulation}\label{section_problem_formulation}
First of all, we present some notations that will be used through out the paper.
Let $\one=[1,\dots,1]^T\in\mathbb{R}^n$, and $I$ be the identity matrix with appropriate dimensions.
Denote $|\cdot|$ as the absolute value of a real number, and $\Null(\cdot)$ as the null space of a matrix.
Given $x=[x_1,\dots,x_n]^T\in\mathbb{R}^n$ and $p\ge1$, the $p$-norm of $x$ is denoted as $\|x\|_p=(\sum_{i=1}^n |x_i|^p)^{1/p}$.
For the sake of simplicity, we omit the subscript when $p=2$, i.e., denoting $\|\cdot\|$ as the $2$-norm.
Let $[\cdot]_{ij}$ be the entry at the $i$th row and $j$th column of a matrix, and $[\cdot]_i$ be the $i$th entry of a vector.
For a symmetric positive (semi-) definite matrix $A$, its positive real eigenvalues are denoted as $0\le\lambda_1(A)\le\lambda_2(A)\le\dots\le\lambda_n(A)$.

Consider $n$ ($n \ge 3$) agents in $\mathbb{R}^2$.
The information topology of the agents is described by a graph $\mathcal{G}=(\mathcal{V},\mathcal{E})$, which consists of a vertex set $\mathcal{V}=\{1,\dots,n\}$ and an edge set $\mathcal{E}\subseteq \mathcal{V} \times \mathcal{V}$.
Each vertex in $\mathcal{G}$ corresponds to an agent.
If $(i,j)\in \mathcal{E}$, then $i$ and $j$ are called to be adjacent, and $i$ receives information from $j$.
The set of neighbors of vertex $i$ is denoted as $\mathcal{N}_i=\{j \in \mathcal{V}: \ (i,j)\in \mathcal{E}\}$.
A graph is undirected if each $(i,j)\in \mathcal{E}$ implies $(j,i)\in \mathcal{E}$, otherwise it is directed.
A path from $i$ to $j$ in a graph is a sequence of distinct nodes starting with $i$
and ending with $j$ such that consecutive vertices are adjacent.
If there is a path between any two vertices of a graph $\mathcal{G}$, then $\mathcal{G}$ is said to be connected.
An undirected circular graph is a connected graph where every vertex has exactly two neighbors.

An incidence matrix of a directed graph is a matrix $E$ with rows indexed by
edges and columns indexed by vertices
\footnote{In some literature such as \cite{graphbook}, the rows of an incidence matrix are indexed by
vertices and the columns are indexed by edges.}.
Suppose $(j,k)$ is the $i$th edge. Then the entry of $E$ in the $i$th row and $k$th column is $1$, the one in the $i$th row and $j$th column is $-1$, and the others in the $i$th row are zero.
Thus by definition, $E\one=0$.
If the graph is connected, the corresponding $E$ has rank $n-1$ \cite[Theorem 8.3.1]{graphbook}.
Then $\Null(E)=\mathrm{span}\{\one\}$.

\subsection{Angle-constrained Target Formation}
\begin{figure}
  \centering
    \begin{tikzpicture}[scale=0.5]
            \coordinate (zim2) at (11,0);
            \coordinate (zim1) at (8,-1);
            \coordinate (zi)   at (0,0);
            \coordinate (zip1) at (-2,9);
            \coordinate (zip2) at (3,11);
            \coordinate (zip3) at (4,12);
            \draw [densely dotted, thin] (zim2)--(zim1)--(zi)--(zip1)--(zip2)--(zip3);
            \def\unitlength{3cm}
            \draw[->, >=latex, very thick] (zi) -- ($(zi)!\unitlength!(zim1)$) node [below] {$-g_{i-1}$};
            \draw[->, >=latex, very thick] (zi) -- ($(zi)!\unitlength!(zip1)$) node [left] {$g_{i}$};
            \draw[->, >=latex, very thick] (zim1) -- ($(zim1)!\unitlength!(zi)$) node [above] {$g_{i-1}$};
            \draw[->, >=latex, very thick] (zip1) -- ($(zip1)!\unitlength!(zi)$) node [left] {$-g_{i}$};
            \draw[->, >=latex, very thick] (zip1) -- ($(zip1)!\unitlength!(zip2)$) node [above] {$g_{i+1}$};
            \def\angleRadius{1cm}
            \draw[->] let \p1=(zi), \p2=(zim1), \p3=(zip1), \n1={atan2(\x2-\x1,\y2-\y1)}, \n2={atan2(\x3-\x1,\y3-\y1)}  in
                    ($(\p1)!\angleRadius!(\p2)$) arc (\n1:\n2:\angleRadius);
            \draw[] let \p1=(zi), \p2=(zim1), \p3=(zip1), \n1={atan2(\x2-\x1,\y2-\y1)}, \n2={atan2(\x3-\x1,\y3-\y1)} in
                    (\p1)+(\n1/2+\n2/2:\angleRadius) node[above right] {$\theta_i$};
            \draw[->] let \p1=(zip1), \p2=(zi), \p3=(zip2), \n1={atan2(\x2-\x1,\y2-\y1)}, \n2={atan2(\x3-\x1,\y3-\y1)}  in
                    ($(\p1)!\angleRadius!(\p2)$) arc (\n1:\n2:\angleRadius);
            \draw[] let \p1=(zip1), \p2=(zi), \p3=(zip2), \n1={atan2(\x2-\x1,\y2-\y1)}, \n2={atan2(\x3-\x1,\y3-\y1)} in
                    (\p1)+(\n1/2+\n2/2:\angleRadius) node[below right] {$\theta_{i+1}$};
            \def\radius{10pt}
            \draw [fill=white](zi) circle [radius=\radius];
            \draw [fill=white](zim1) circle [radius=\radius];
            \draw [fill=white](zip1) circle [radius=\radius];
            \draw [fill=white](zip2) circle [radius=\radius];
            \draw (zi) node[below=\radius/2] {$z_{i}$};
            \draw (zim1) node[below=\radius/2] {$z_{i-1}$};
            \draw (zip1) node[below left=\radius/2] {$z_{i+1}$};
\end{tikzpicture}
  \caption{Illustration of a circular formation.}
  \label{fig_circular_formation}
\end{figure}
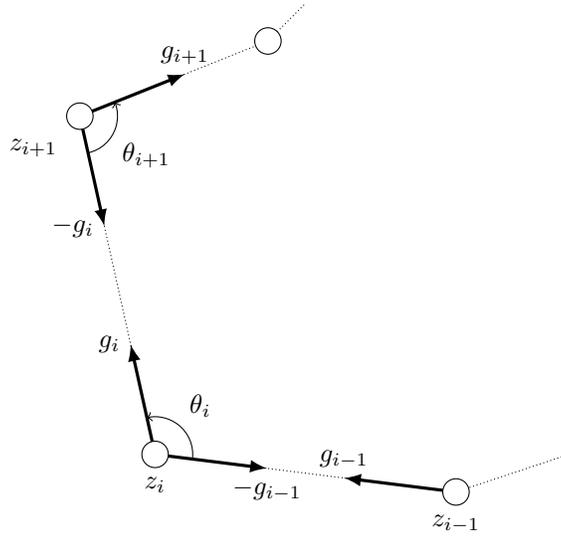

In this paper we consider the distributed control of circular (or polygon) formations.
The underlying information flow among the agents is described by a fixed undirected circular graph.
By indexing the agents properly, we can have $\mathcal{N}_i=\{i+1,i-1\}$ for $i\in\{1,\dots,n\}$,
which means agent $i$ can measure the bearings of agents $i-1$ and $i+1$.
The indices $i+1$ and $i-1$ are taken modulo $n$.
Denote the position of sensor $i$ as $z_i\in\mathbb{R}^2$, and the edge between agents $i$ and $i+1$ as $e_i=z_{i+1}-z_i$.
The unit-length vector $g_i={e_i}/{\|e_i\|}$ characterizes the relative bearing between agents $i+1$ and $i$ (see Fig.~\ref{fig_circular_formation}).
Hence the measurements of agent $i$ consist of $g_{i}$ and $-g_{i-1}$.
It should be noted that agent $i$ may measure $g_{i}$ and $-g_{i-1}$ in its local coordinate frame.
While analyzing the dynamics of the entire system, we need to write these bearing measurements in a global coordinate frame.

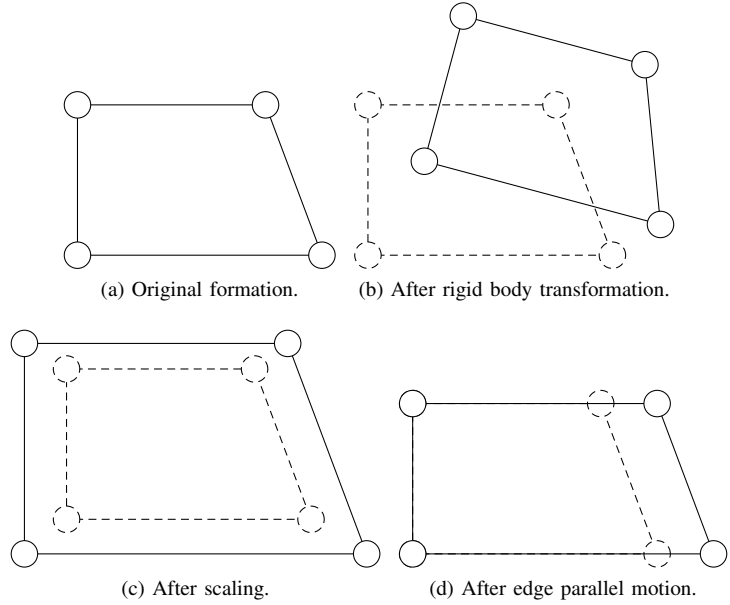
\begin{figure}
    \centering
    \def\myScale{0.35}
    \subfloat[Original formation.]
{
\begin{tikzpicture}[scale=0.5]
    \coordinate (z1) at (0,0);
    \coordinate (z2) at (0,4);
    \coordinate (z3) at (5,4);
    \coordinate (z4) at (6.5,0);
    \draw (z1)--(z2)--(z3)--(z4)--cycle;
    \def\radius{10pt}
    \draw [fill=white](z1) circle [radius=\radius];
    \draw [fill=white](z2) circle [radius=\radius];
    \draw [fill=white](z3) circle [radius=\radius];
    \draw [fill=white](z4) circle [radius=\radius];
\end{tikzpicture}
}
\subfloat[After rigid body transformation.]
{
\begin{tikzpicture}[scale=0.5]
    \coordinate (z1) at (0,0);
    \coordinate (z2) at (0,4);
    \coordinate (z3) at (5,4);
    \coordinate (z4) at (6.5,0);
    \draw[densely dashed] (z1)--(z2)--(z3)--(z4)--cycle;
    \def\radius{10pt}
    \draw [fill=white,densely dashed](z1) circle [radius=\radius];
    \draw [fill=white,densely dashed](z2) circle [radius=\radius];
    \draw [fill=white,densely dashed](z3) circle [radius=\radius];
    \draw [fill=white,densely dashed](z4) circle [radius=\radius];

    \begin{scope}[shift={(1.5,2.5)},rotate=-15]
    \coordinate (z1) at (0,0);
    \coordinate (z2) at (0,4);
    \coordinate (z3) at (5,4);
    \coordinate (z4) at (6.5,0);
    \end{scope}
    \draw (z1)--(z2)--(z3)--(z4)--cycle;
    \def\radius{10pt}
    \draw [fill=white](z1) circle [radius=\radius];
    \draw [fill=white](z2) circle [radius=\radius];
    \draw [fill=white](z3) circle [radius=\radius];
    \draw [fill=white](z4) circle [radius=\radius];
\end{tikzpicture}
}
\\
\subfloat[After scaling.]
{
\begin{tikzpicture}[scale=0.5]
    \coordinate (z1) at (0,0);
    \coordinate (z2) at (0,4);
    \coordinate (z3) at (5,4);
    \coordinate (z4) at (6.5,0);
    \draw[densely dashed] (z1)--(z2)--(z3)--(z4)--cycle;
    \def\radius{10pt}
    \draw [fill=white,densely dashed](z1) circle [radius=\radius];
    \draw [fill=white,densely dashed](z2) circle [radius=\radius];
    \draw [fill=white,densely dashed](z3) circle [radius=\radius];
    \draw [fill=white,densely dashed](z4) circle [radius=\radius];

    \begin{scope}[scale around={1.4:(2.8,2.3)}] 
    \coordinate (z1) at (0,0);
    \coordinate (z2) at (0,4);
    \coordinate (z3) at (5,4);
    \coordinate (z4) at (6.5,0);
    \end{scope}
    \draw (z1)--(z2)--(z3)--(z4)--cycle;
    \def\radius{10pt}
    \draw [fill=white](z1) circle [radius=\radius];
    \draw [fill=white](z2) circle [radius=\radius];
    \draw [fill=white](z3) circle [radius=\radius];
    \draw [fill=white](z4) circle [radius=\radius];
\end{tikzpicture}
}
\subfloat[After edge parallel motion.]
{
\begin{tikzpicture}[scale=0.5]
    \coordinate (z1) at (0,0);
    \coordinate (z2) at (0,4);
    \coordinate (z3) at (5,4);
    \coordinate (z4) at (6.5,0);
    \draw[densely dashed] (z1)--(z2)--(z3)--(z4)--cycle;
    \def\radius{10pt}
    \draw [fill=white,densely dashed](z1) circle [radius=\radius];
    \draw [fill=white,densely dashed](z2) circle [radius=\radius];
    \draw [fill=white,densely dashed](z3) circle [radius=\radius];
    \draw [fill=white,densely dashed](z4) circle [radius=\radius];

    \coordinate (z1) at (0,0);
    \coordinate (z2) at (0,4);
    \begin{scope}[xshift=1.5cm]
    \coordinate (z3) at (5,4);
    \coordinate (z4) at (6.5,0);
    \end{scope}
    \draw (z1)--(z2)--(z3)--(z4)--cycle;
    \def\radius{10pt}
    \draw [fill=white](z1) circle [radius=\radius];
    \draw [fill=white](z2) circle [radius=\radius];
    \draw [fill=white](z3) circle [radius=\radius];
    \draw [fill=white](z4) circle [radius=\radius];
\end{tikzpicture}
}
    \caption{The angle-constrained target formation is invariant to four motion patterns. (a) Original formation; (b) after rigid body transformation; (c) after scaling; (d) after edge parallel motion.}
    \label{fig_bearing_constained_formation}
\end{figure}

The angle subtended by agents $i+1$ and $i-1$ at agent $i$ is denoted as $\theta_i\in[0,2\pi)$.
More specifically, rotating $-g_{i-1}$ counterclockwise through an angle $\theta_i$ about agent $i$ yields $g_i$ (see Fig.~\ref{fig_circular_formation}).
Hence $\theta_i$ and $\theta_{i+1}$ are on the same side of edge $e_i$ for all $i\in\{1,\dots,n\}$.
Consequently the quantity $\sum_{i=1}^n \theta_i$ is invariant to the positions of the agents because the sum of the interior angles of a polygon is constant.
In the target formation, the angle $\theta_i$ is specified as $\theta_i^*\in[0,2\pi)$.
The target angles $\{\theta_i^*\}_{i=1}^n$ should be feasible such that there exist $\{z_i\}_{i=1}^n$ ($z_i\ne z_j$ for $i\ne j$) to realize the target formation.
Because the target formation is only constrained by angles, the realization will be non-unique.
This paper makes no assumptions about parallel rigidity \cite{eren2003,bishopconf2011rigid,ErenIJC} of the formation.
As shown in Fig.~\ref{fig_bearing_constained_formation}, the angle-constrained target formation is \emph{invariant} to the following motion patterns: rigid body transformation, scaling and edge parallel motion.
The term invariant as used here means that these motions will not change the angles in the formation.
Moreover, it is notable that $\sum_{i=1}^n\theta_i\equiv\sum_{i=1}^n\theta_i^*$ regardless the motion of agents.


\subsection{Proposed Control Law}
The control task can be summarized as this: steer agents from their initial positions $\{z_i(0)\}_{i=1}^n$ to final positions such that $\theta_i=\theta_i^*$ for all $i\in\{1,\dots,n\}$.
For a proper definition of $\theta_i$, we need to assume $z_i(0)\ne z_j(0)$ for all $i\ne j$.
The feedback angle error of agent $i$ is defined as
\begin{align}
    \varepsilon_i
    &=\cos\theta_i-\cos\theta_i^* \\ \nonumber
    &=-g_i^Tg_{i-1}-\cos\theta_i^*.
\end{align}
The reason why we use cosine functions to define the angle error $\varepsilon_i$ is that $\cos\theta_i$ can be conveniently expressed as the inner product of the two bearing measurements $g_i$ and $-g_{i-1}$.
Suppose the motion model of each agent as a single integrator.
The proposed control law for agent $i$ is
\begin{align}\label{eq_controlLaw}
        \dot{z}_i=\sgn(\varepsilon_i)|\varepsilon_i|^a(g_i-g_{i-1}),
\end{align}
where $a\in(0,1]$ and $\sgn(\cdot)$ is the sign function defined by
\begin{align}
        \sgn(\varepsilon_i)=\left\{
          \begin{array}{l l}
            1  & \mbox{if } \varepsilon_i>0\\
            0  & \mbox{if } \varepsilon_i=0\\
            -1 & \mbox{if } \varepsilon_i<0
          \end{array}.\right.
\end{align}
In the special case when $a=1$, control law \eqref{eq_controlLaw} becomes $\dot{z}_i=\varepsilon_i(g_i-g_{i-1})$ because $\sgn(\varepsilon_i)|\varepsilon_i|^a=\varepsilon_i$.

\begin{my_remark}
    As will be shown later, control law \eqref{eq_controlLaw} ensures local exponential stability if $a=1$, and local finite-time stability if $a\in(0,1)$.
    Loosely speaking, finite-time stability means $\varepsilon_i$ for all $i$ converges to zero in finite time.
    See \cite{bernstein2000} or \cite[Section 4.6]{lyapunovbook} for a formal definition of finite-time stability of nonlinear systems.
    In this paper, we will present a unified proof of the exponential and finite-time stability based on Lyapunov approaches.
\end{my_remark}

\begin{my_remark}
    If $a>0$, it is straightforward to see $\sgn(\varepsilon_i)|\varepsilon_i|^a$ is continuous in $\varepsilon_i$.
    If $a=0$, control law \eqref{eq_controlLaw} is discontinuous in $\varepsilon_i$.
    The discontinuous case with $a=0$ is out of the scope of this paper.
    One may refer to \cite{corte2006automatica,corte2008CSM} for finite-time stability of discontinuous dynamic systems.
\end{my_remark}

Our proposed control law is inspired by the one in \cite{bishopSCL}, where the control law steers agents moving along the bisectors of the angles in a triangle.
The velocity direction in control law \eqref{eq_controlLaw} is also along the bisector of $\theta_i$.
But the feedback angle error and velocity are differently defined in \eqref{eq_controlLaw} compared to \cite{bishopSCL}.
It is noticed that $g_i-g_{i-1}$ will vanish when $\theta_i=\pi$.
Hence the control law is ineffective in the case of $\theta_i=\pi$ even though $\varepsilon_i$ is still nonzero.
Moreover, when $\theta_i=0$, agents $i-1$ and $i+1$ are on the same side of agent $i$.
Since bearing information is usually measured by optical sensors such as cameras, the bearing of agent $i-1$ or $i+1$ may not measurable by agent $i$ due to line-of-sight occlusion in the case of $\theta_i=0$.
Therefore, we make the following assumption.

\begin{my_assumption}\label{assumption}
    In the target formation, $\theta_i^*\ne 0$ and $\theta_i^*\ne \pi$ for all $i\in\{1,\dots,n\}$.
\end{my_assumption}

By Assumption \ref{assumption}, the angle $\theta_i^*$ is in either $(0,\pi)$ or $(\pi,2\pi)$.
In other words, no three consecutive agents in the target formation are collinear.
The collinear case is a difficulty in many formation control problems (see \cite{Francis2009IJC,Francis2010TAC,Huang2012,bishopconf2011quad} for example).
Because the angle error $\varepsilon_i$ is defined using cosine functions and $g_i-g_{i-1}$ vanishes in the case of $\theta_i=\pi$,
control law \eqref{eq_controlLaw} inherently is not able to ensure global convergence to arbitrary feasible target formations.
However, the benefit of control law \eqref{eq_controlLaw} is that its dynamics can be conveniently analyzed based on Lyapunov approaches.
At last, we would like to emphasize that although $g_i$ and $g_{i-1}$ in \eqref{eq_controlLaw} are expressed in a global coordinate frame, the control law can be implemented using agent $i$'s bearing-only measurements in its local coordinate frame.

\section{Preliminary Results}\label{section_useful_lemma}
In order to analyze the stability by the proposed control law, we need to prove and introduce the following results.

\begin{my_lemma}\label{lemma_infimum_angle}
    Let $A\in\mathbb{R}^{n\times n}$ be a positive semi-definite matrix with $\lambda_1(A)=0$ and $\lambda_2(A)>0$.
    An eigenvector with the zero eigenvalue is $\one=[1,\dots,1]^T\in\mathbb{R}^n$.
    Let
    \begin{align}
        \mbox{
        $\mathcal{U}=\{x\in\mathbb{R}^n: \|x\|=1$ and nonzero entries of $x$ are not with the same sign\}.
        }
    \end{align}
    Then
    \begin{align}\label{eq_xAx_inf}
        \inf_{x\in\mathcal{U}} x^TAx = \frac{\lambda_2(A)}{n}.
    \end{align}
\end{my_lemma}
\begin{proof}
By orthogonally projecting $x$ to $\one$ and the orthogonal complement of $\one$, we can decompose $x$ as
\begin{align}
    x=c_0 x_{0}+ c_1 x_{1},
\end{align}
where $\|x_0\|=\|x_1\|=1$, $x_{0} \parallel \one$, $x_{1}\perp \one$, $c_0=x_0^Tx$ and $c_1=x_1^Tx$.
Because $A\one=0$, $\one^TA=0$ and $x_{0} \parallel \one$, we have
\begin{align}\label{eq_xAx}
    x^TAx=c_1^2x_1^TAx_1\ge c_1^2\lambda_2(A),
\end{align}
where the last inequality uses the well-known fact
\begin{align}
    \min_{x_1\perp\one,\|x_1\|=1}x_1^TAx_1=\lambda_2(A).
\end{align}
Denote $\varphi\in[0,\pi]$ as the angle between $\one$ and $x$.
Then $c_1=\sin\varphi$.
Geometrically speaking, the vector $\one$ is isolated from $x\in\mathcal{U}$ by the hyper-planes $[x]_i=0$ with $i\in\{1,\dots,n\}$.
Let $p_i\in\mathbb{R}^n$ be a vector with zero as the $i$th entry and one as the others.
Then $p_i$ is the orthogonal projection of $\one$ onto the hyper-plane $[x]_i=0$.
Note the set $\mathcal{U}$ is open.
The infimum angle $\varphi_{\inf}$ between $\one$ and $x$ is the angle between $\one$ and $p_i$.
Thus $\one^Tp_i=\|\one\|\|p_i\|\cos\varphi_{\inf}$, which implies $\cos\varphi_{\inf}=\sqrt{n-1}/\sqrt{n}$ and $\sin\varphi_{\inf}=1/\sqrt{n}$.
The supremum angle $\varphi_{\sup}$ between $\one$ and $x$ is the angle between $\one$ and $-p_i$. In that case, we have $\cos\varphi_{\sup}=-\sqrt{n-1}/\sqrt{n}$ and $\sin\varphi_{\sup}=1/\sqrt{n}$.
Therefore,
\begin{align}
    \inf\{x\in\mathcal{U}: c_1=\sin\varphi\}=\frac{1}{\sqrt{n}},
\end{align}
substituting which into $\eqref{eq_xAx}$ yields $\eqref{eq_xAx_inf}$.
\end{proof}

\begin{my_lemma}\label{lemma_gammaupperbound}
    Let $x(t)$ be a real positive scalar variable for all $t\in[0,+\infty)$. Given $k\in(0,1)$, if the time derivative of $x(t)$ satisfies
    \begin{align}\label{eq_x_dot}
        |\dot{x}(t)| \le \exp\left(\int_{0}^t -\frac{k}{x(\tau)}\mathrm{d} \tau\right), \quad t\in[0,+\infty),
    \end{align}
    then $x(t)$ for all $t\in[0,+\infty)$ has a finite upper bound.
\end{my_lemma}

\begin{proof}
We will find an explicit finite upper bound by repeatedly calculating the integral in \eqref{eq_x_dot} and using the fact $x(t)\le x(0)+\int_{0}^t |\dot{x}(\tau)|\mathrm{d}\tau$.

First of all, because $x>0$, we have $-k/x<0$ and hence
\begin{align}
    |\dot{x}(t)|
    \le \exp(0)=1.
\end{align}
Then
\begin{align}
    x(t)
    \le x(0)+\int_0^t1\mathrm{d}\tau\le t+c, \quad c\ge x(0).
\end{align}
Substituting the above inequality back into \eqref{eq_x_dot} gives
\begin{align}
    |\dot{x}(t)| \nonumber
    &\le \exp\left(\int_{0}^t -\frac{k}{\tau+c}\mathrm{d} \tau\right) \\ \nonumber
    &= \exp\left(-k \ln\frac{t+c}{c}\right) \\
    &= \left(\frac{c}{t+c}\right)^k.
\end{align}
Then
\begin{align}
    x(t) \nonumber
    &\le x(0)+\int_0^t  \left(\frac{c}{\tau+c}\right)^k \mathrm{d} \tau \\ \nonumber
    &=x(0)+\frac{c^k}{1-k}\left[(t+c)^{1-k}-c^{1-k}\right] \\
    &<\frac{c^k}{1-k}(t+c)^{1-k},
\end{align}
where the last inequality uses the fact $c\ge x(0)$, $1-k<1$ and hence $x(0)-c/(1-k)<0$.
Again substituting the above inequality into \eqref{eq_x_dot} gives
\begin{align}
    |\dot{x}(t)| \nonumber
    &< \exp\left(\int_{0}^t -\frac{(1-k)k}{c^k}(\tau+c)^{k-1} \mathrm{d} \tau\right) \\ \nonumber
    &=\exp\left( -\frac{1-k}{c^k} \left((t+c)^k-c^k\right)\right) \\
    &=e^{1-k}\exp\left( -\frac{1-k}{c^k}(t+c)^k\right).
\end{align}
Denote $\mu=(1-k)/c^k$. Then for all $t\in[0,+\infty)$
\begin{align}
    x(t) \nonumber
    &\le x(0)+e^{1-k}\int_0^{t}e^ {-\mu(\tau+c)^k} \mathrm{d}\tau \\
    &\le x(0)+e^{1-k}\int_0^{+\infty}e^ {-\mu(\tau+c)^k} \mathrm{d}\tau.
\end{align}
Let $s=\mu(\tau+c)^k$ and hence $\mathrm{d}\tau=(1/k)\mu^{-1/k}s^{1/k-1}\mathrm{d}s$.
Then above integral becomes
\begin{align}
    \int_0^{+\infty}e^{-\mu(\tau+c)^k} \mathrm{d}\tau \nonumber
    &=\frac{1}{k}\mu^{-\frac{1}{k}} \int_{\mu c^k}^{+\infty} e^{-s}s^{\frac{1}{k}-1} \mathrm{d}s \\
    &<\frac{1}{k}\mu^{-\frac{1}{k}} \int_{0}^{+\infty} e^{-s}s^{\frac{1}{k}-1} \mathrm{d}s.
\end{align}
For any positive real constant $\alpha$, the integral $\Gamma(\alpha)=\int_{0}^{+\infty} e^{-s}s^{\alpha-1} \mathrm{d}s$ in fact is the Gamma function of $\alpha$ and has a finite value.
Therefore, we conclude that $x(t)$ for all $t\in[0,+\infty)$ has a finite upper bound
\begin{align}
    x(t)<x(0)+\frac{1}{k}\mu^{-\frac{1}{k}} e^{1-k}\Gamma\left(\frac{1}{k}\right).
\end{align}
\end{proof}


\begin{my_lemma}[{\cite[Lemma 2]{wanglongarxiv}}]\label{lemma_wanglong_lemma}
    Let $x_1,\dots,x_n\ge0$. Given $p\in(0,1]$, then
    \begin{align}
        \left(\sum_{i=1}^n x_i\right)^p  \le \sum_{i=1}^n x_i^p  \le  n^{1-p}\left(\sum_{i=1}^n x_i\right)^p.
    \end{align}
\end{my_lemma}

\begin{my_lemma}[{\cite[Corollary 5.4.5]{bookmatrix}}]\label{lemma_vector_norm_equivalent}
    Let $\|\cdot\|_{\alpha}$ and $\|\cdot\|_{\beta}$ be any two vector norms on $\mathbb{R}^n$. Then there exist finite positive constants $C_m$ and $C_M$ such that $C_m\|x\|_{\alpha}\le \|x\|_{\beta}\le C_M \|x\|_{\alpha}$ for all $x\in\mathbb{R}^n$.
\end{my_lemma}


\section{Stability Analysis}\label{section_stability}
In this section, we present a unified way to prove that the target formation by control law \eqref{eq_controlLaw} is locally exponentially stable if $a=1$, and locally finite-time stable if $a\in(0,1)$.

\subsection{Basic Stability Analysis}
Denote $\varepsilon=[\varepsilon_1,\dots,\varepsilon_n]^T\in\mathbb{R}^n$ and $z=[z_1^T,\dots,z_n^T]^T\in\mathbb{R}^{2n}$.
From \eqref{eq_controlLaw}, it is straightforward to see $\varepsilon=0$ implies $\dot{z}=0$ and then $\dot{\varepsilon}=0$.
Hence $\varepsilon=0$ is an equilibrium of the $\varepsilon$-dynamics.
We here omit giving the dynamic equation of $\varepsilon$, whose dynamics, however, will be directly used in the following analysis.
Now consider the Lyapunov candidate
\begin{align}
    V = \frac{1}{a+1}\sum_{i=1}^n |\varepsilon_i|^{a+1}.
\end{align}
Then $V$ can also be written as $V=1/(a+1)\|\varepsilon\|_{a+1}^{a+1}$ where $\|\cdot\|_{a+1}$ denotes the $(a+1)$-norm.
Clearly $V$ is positive definite with respect to $\varepsilon=0$.
In the case of $a=1$, the Lyapunov candidate becomes $V =1/2\varepsilon^T\varepsilon$, which is an ordinary quadratic function of $\varepsilon$.

If $\varepsilon_i>0$, $\frac{\partial |\varepsilon_i|^{a+1}}{\partial \varepsilon_i}=\frac{\partial \varepsilon_i^{a+1}}{\partial \varepsilon_i}=(a+1)\varepsilon_i^a=(a+1)\sgn(\varepsilon_i)|\varepsilon_i|^a$ and $\lim_{\varepsilon_i\rightarrow 0^+}\frac{\partial |\varepsilon_i|^{a+1}}{\partial \varepsilon_i}=0$;
if $\varepsilon_i<0$, $\frac{\partial |\varepsilon_i|^{a+1}}{\partial \varepsilon_i}=\frac{\partial (-\varepsilon_i)^{a+1}}{\partial \varepsilon_i}=-(a+1)(-\varepsilon_i)^a=(a+1)\sgn(\varepsilon_i)|\varepsilon_i|^a$ and $\lim_{\varepsilon_i\rightarrow 0^-}\frac{\partial |\varepsilon_i|^{a+1}}{\partial \varepsilon_i}=0$. Therefore, for all $\varepsilon_i$ we have
\begin{align}
    \frac{\partial |\varepsilon_i|^{a+1}}{\partial \varepsilon_i}=(a+1)\sgn(\varepsilon_i)|\varepsilon_i|^a.
\end{align}
For the sake of simplicity, denote $\sigma_i=\sgn(\varepsilon_i)|\varepsilon_i|^a$ and $\sigma=[\sigma_1,\dots,\sigma_n]^T\in\mathbb{R}^n$.
The time derivative of $V $ is
\begin{align}\label{eq_V_dot}
    \dot{V}  \nonumber
    &=\frac{1}{a+1}\sum_{i=1}^n \frac{\partial |\varepsilon_i|^{a+1}}{\partial \varepsilon_i}\dot{\varepsilon}_i \\ \nonumber
    &=\sum_{i=1}^n \sigma_i \dot{\varepsilon}_i \\ \nonumber
    &=\sum_{i=1}^n \sigma_i (-g_i^T\dot{g}_{i-1}-g_{i-1}^T\dot{g}_i) \\ \nonumber
    &=\sum_{i=1}^n \sigma_i (-g_i^T\dot{g}_{i-1}) +
        \sum_{i=1}^n \sigma_i(-g_{i-1}^T\dot{g}_i) \\ \nonumber
    &=\sum_{i=1}^n \sigma_{i+1} (-g_{i+1}^T\dot{g}_{i}) +
            \sum_{i=1}^n \sigma_i(-g_{i-1}^T\dot{g}_i) \\
    &=-\sum_{i=1}^n (\sigma_{i+1}g_{i+1} +\sigma_ig_{i-1})^T\dot{g}_i.
\end{align}
Because $g_i=e_i/\|e_i\|$, we have
\begin{align}\label{eq_g_dot}
    \dot{g}_i
    &=\frac{1}{\|e_i\|}P_i\dot{e}_i,
\end{align}
where $P_i=I-g_ig_i^T$.
Note $P_i$ is an orthogonal projection matrix satisfying $P_i^T=P_i$ and $P_i^2=P_i$.
Moreover, $P_i$ is positive semi-definite and $\Null(P_i)=\mathrm{span}\{g_i\}$.
From control law \eqref{eq_controlLaw}, we have
\begin{align}
    \dot{z}_i       &= \sigma_i(g_i-g_{i-1}), \\
    \dot{z}_{i+1}   &= \sigma_{i+1}(g_{i+1}-g_i),
\end{align}
and hence
\begin{align}\label{eq_e_dot}
    \dot{e}_i \nonumber
    &=\dot{z}_{i+1}-\dot{z}_{i} \\
    &=\sigma_{i+1}g_{i+1}+\sigma_ig_{i-1}  -(\sigma_{i+1}+\sigma_i) g_i.
\end{align}
Because $P_ig_i=0$, substituting the above $\dot{e}_i$ into \eqref{eq_g_dot} gives
\begin{align}
    \dot{g}_i \nonumber
    &=\frac{1}{\|e_i\|}P_i(\sigma_{i+1}g_{i+1}+\sigma_ig_{i-1}  -(\sigma_{i+1}+\sigma_i) g_i) \\
    &=\frac{1}{\|e_i\|}P_i(\sigma_{i+1}g_{i+1}+\sigma_ig_{i-1}).
\end{align}
Substituting the above $\dot{g}_i$ into \eqref{eq_V_dot} yields
\begin{align}\label{eq_V_dot_2}
    \dot{V}
    =-\sum_{i=1}^n \frac{1}{\|e_i\|}(\sigma_{i+1}g_{i+1} +\sigma_ig_{i-1})^TP_i(\sigma_{i+1}g_{i+1}+\sigma_ig_{i-1})\le0.
\end{align}
Now we can claim the equilibrium $\varepsilon=0$ is at least Lyapunov stable.

\subsection{Exponential and Finite-time Stability Analysis}
In order to further prove the exponential and finite-time stability, we need the following results.

For an arbitrary angle $\alpha$, the rotation matrix
\begin{align}
    R(\alpha)=\left[
                \begin{array}{cc}
                  \cos \alpha & -\sin\alpha \\
                  \sin\alpha & \cos\alpha \\
                \end{array}
              \right]\in\mathbb{R}^{2\times2}
\end{align}
rotates a vector in $\mathbb{R}^2$ counterclockwise through an angle $\alpha$ about the origin.
Thus it is clear that for all nonzero $x\in\mathbb{R}^2$, $x^TR(\alpha)x>0$ when $\alpha\in(-\pi/2,\pi/2)$ (mod $2\pi$); $x^TR(\alpha)x=0$ when $\alpha=\pm\pi/2$ (mod $2\pi$); and $x^TR(\alpha)x<0$ when $\alpha\in(\pi/2,3\pi/2)$ (mod $2\pi$).
Moreover, $R^{-1}(\alpha)=R^T(\alpha)=R(-\alpha)$ and $R(\alpha_1)R(\alpha_2)=R(\alpha_1+\alpha_2)$ for arbitrary angles $\alpha_1$ and $\alpha_2$.

\begin{my_lemma}\label{lemma_gi_perp}
Let $g_i^{\perp}=R(\pi/2)g_i$.
It is obvious that $\|g_i^{\perp}\|=1$ and $(g_i^{\perp})^Tg_i=0$.
Furthermore,
\begin{enumerate} [(i)]
  \item $ P_i=g_i^{\perp}(g_i^{\perp})^T$.
  \item For $i\ne j$, $(g_i^{\perp})^Tg_j=-(g_j^{\perp})^T g_i$.
  \item $(g_i^{\perp})^Tg_{i-1}>0$ if $\theta_i\in(0,\pi)$; and $(g_i^{\perp})^Tg_{i-1}<0$ if $\theta_i\in(\pi,2\pi)$.
\end{enumerate}
\end{my_lemma}
\begin{proof}
(i) Denote $G_i=[g_i,g_i^{\perp}]\in\mathbb{R}^{2\times 2}$. Then $G_i$ is an orthogonal matrix satisfying $G_i^TG_i=I$.
Hence we have
\begin{align}
    g_ig_i^T+g_i^{\perp}(g_i^{\perp})^T=G_iG_i^T=I.
\end{align}
Thus $g_i^{\perp}(g_i^{\perp})^T=I-g_ig_i^T=P_i$.

(ii) $(g_i^{\perp})^Tg_j=g_i^TR^T(\pi/2)g_j=g_i^TR(-\pi/2)g_j=g_i^TR(-\pi)R(\pi/2)g_j=g_i^TR(-\pi)g_j^{\perp}=g_i^T(-I)g_j^{\perp}= -(g_j^{\perp})^T g_i$.

(iii)
By the definition of $\theta_i$, we have $g_i=R(\theta_i)(-g_{i-1})$ and hence $g_{i-1}=-R(-\theta_i)g_i$.
Then
\begin{align}
    (g_i^{\perp})^Tg_{i-1}
    =-g_i^TR\left(-\frac{\pi}{2}\right)R(-\theta_i)g_i
    =-g_i^TR\left(-\frac{\pi}{2}-\theta_i\right)g_i
    =-\cos\left(-\frac{\pi}{2}-\theta_i\right)
    =\sin\theta_i.
\end{align}
Then it is straightforward to see the result in (iii).
\end{proof}


Because $P_i=g_i^{\perp}(g_i^{\perp})^T$ as shown in Lemma \ref{lemma_gi_perp} (i), $\dot{V}$ in \eqref{eq_V_dot_2} becomes
\begin{align}\label{eq_V_dot_3}
    \dot{V} \nonumber
    &=-\sum_{i=1}^n \frac{1}{\|e_i\|}  \left((g_i^{\perp})^T(\sigma_{i+1}g_{i+1}+\sigma_ig_{i-1}) \right)^2 \\ \nonumber
    &\le -\frac{1}{\sum_{i=1}^n\|e_i\|}  \sum_{i=1}^n \left(\sigma_{i+1}(g_i^{\perp})^Tg_{i+1}+\sigma_i(g_i^{\perp})^Tg_{i-1}\right)^2 \\
    &= -\frac{1}{\sum_{i=1}^n\|e_i\|}
    \left\|\left[
      \begin{array}{c}
        \sigma_{2}(g_1^{\perp})^Tg_{2}+\sigma_1(g_1^{\perp})^Tg_{n} \\
        \vdots \\
        \sigma_{1}(g_n^{\perp})^Tg_{1}+\sigma_n(g_n^{\perp})^Tg_{n-1} \\
      \end{array}
    \right]\right\|^2.
\end{align}
Because
\begin{align}
\nonumber &\left[
      \begin{array}{c}
        \sigma_{2}(g_1^{\perp})^Tg_{2}+\sigma_1(g_1^{\perp})^Tg_{n} \\
        \vdots \\
        \sigma_{1}(g_n^{\perp})^Tg_{1}+\sigma_n(g_n^{\perp})^Tg_{n-1} \\
      \end{array}
    \right]\\ \nonumber
    &=\left[
          \begin{array}{ccccc}
            (g_1^{\perp})^Tg_n & (g_1^{\perp})^Tg_2 & 0 & \dots & 0 \\
            0                  & (g_2^{\perp})^Tg_1 & (g_2^{\perp})^Tg_3 & \dots & 0 \\
            0                  & 0                  & (g_3^{\perp})^Tg_2 & \dots & 0 \\
            \vdots             & \vdots & \vdots & \ddots & \vdots \\
            (g_n^{\perp})^Tg_1 & 0 & \dots & 0 & (g_n^{\perp})^Tg_{n-1} \\
          \end{array}
        \right]
        \left[
          \begin{array}{c}
            \sigma_1 \\
            \sigma_2 \\
            \sigma_3 \\
            \vdots \\
            \sigma_n \\
          \end{array}
        \right] \\
    &=\underbrace{\left[
          \begin{array}{ccccc}
            1      & -1     & 0      & \dots & 0 \\
            0      & 1      & -1     & \dots & 0 \\
            0      & 0      & 1      & \dots & 0 \\
            \vdots & \vdots & \vdots & \ddots & \vdots \\
            -1     & 0      & \dots  & 0 & 1 \\
          \end{array}
        \right]}_{E\in\mathbb{R}^{n\times n}}
        \underbrace{\left[
          \begin{array}{ccccc}
            (g_1^{\perp})^Tg_n & 0 & 0 & \dots & 0 \\
            0                  & (g_2^{\perp})^Tg_1 & 0 & \dots & 0 \\
            0                  & 0 & (g_3^{\perp})^Tg_2 & \dots & 0 \\
            \vdots & \vdots & \vdots & \ddots & \vdots \\
            0 & 0 & \dots & 0 & (g_n^{\perp})^Tg_{n-1} \\
          \end{array}
        \right]}_{D\in\mathbb{R}^{n\times n}}
        \underbrace{
        \left[
          \begin{array}{c}
            \sigma_1 \\
            \sigma_2 \\
            \sigma_3 \\
            \vdots \\
            \sigma_n \\
          \end{array}
        \right]}_{\sigma\in\mathbb{R}^n},
\end{align}
where the last equality uses the fact that $(g_i^{\perp})^Tg_{i-1}=-(g_{i-1}^{\perp})^Tg_{i}$ as shown in Lemma \ref{lemma_gi_perp} (ii), \eqref{eq_V_dot_3} can be rewritten as
\begin{align}\label{eq_V_dot_4}
    \dot{V}\le -\frac{1}{\sum_{i=1}^n\|e_i\|} \sigma^T D^TE^TED \sigma.
\end{align}

We now are ready to present the main results of this paper.
\begin{my_theorem}\label{theorem_stability}
    Under Assumption \ref{assumption}, the equilibrium $\varepsilon=0$ is locally exponentially stable by control law \eqref{eq_controlLaw} if $a=1$, and locally finite-time stable if $a\in(0,1)$.
\end{my_theorem}
\begin{proof}
The proof consists of two steps. Step 1: prove $\sigma^TD^TE^TED\sigma$ is lower-bounded by $K V ^{\frac{2a}{a+1}}$ with $K $ as a positive constant. Step 2: prove $1/\sum_{i=1}^n \|e_i\|$ is bounded from below by a positive constant.

\emph{Step 1:}

Suppose $\varepsilon\ne0 \Leftrightarrow\sigma\ne0$. Write
\begin{align}\label{eq_threeiterms}
    \sigma^TD^TE^TED\sigma
    =\frac{\sigma^TD^TE^TED\sigma}{\sigma^TD^TD\sigma}  \frac{\sigma^TD^TD\sigma}{V ^{\frac{2a}{a+1}}}  V ^{\frac{2a}{a+1}}.
\end{align}

First, denote $\Omega(c)=\{\varepsilon:\ V(\varepsilon)\le c\}$ with $c>0$ as the level set of $V(\varepsilon)$. Since $\dot{V}\le0$, the level set $\Omega(c)$ is positively invariant with respect to \eqref{eq_controlLaw}.
At the equilibrium point $\varepsilon=0$ (i.e., $\theta_i=\theta_i^*$ for all $i\in\{1,\dots,n\}$), we have $[D]_{ii}=(g_i^{\perp})^Tg_{i-1}\ne 0$ because $\theta_i^*\ne 0$ or $\pi$ as stated in Assumption \ref{assumption}.
Thus by continuity there exists a sufficiently small $c$ such that
$[D]_{ii}\ne 0$ for every point in $\Omega(c)$.
Then $D^TD=D^2$ is positive definite and hence $\lambda_1(D^TD)>0$ for all $\varepsilon\in\Omega(c)$.
Because $V=1/(a+1)\|\varepsilon\|_{a+1}^{a+1}$, the set $\Omega(c)$ is compact given a sufficiently small $c$.
Hence there exists a lower bound $\underline{\lambda}_1(D^TD)>0$ such that $\lambda_1(D^TD)\ge\underline{\lambda}_1(D^TD)$ and consequently
\begin{align}\label{eq_lambda1_DD_bound}
    \sigma^TD^TD\sigma \ge \underline{\lambda}_1(D^TD)\sigma^T\sigma,
\end{align}
for all $\varepsilon\in\Omega(c)$.
In addition, since $2a/(a+1)\in(0,1]$, we have
\begin{align}\label{eq_V_2aa1_bound}
    V ^{\frac{2a}{a+1}} \nonumber
    &=\left(\frac{1}{a+1}\right)^{\frac{2a}{a+1}} \left(\sum_{i=1}^n |\varepsilon_i|^{a+1}\right)^{\frac{2a}{a+1}} \\ \nonumber
    &\le \left(\frac{1}{a+1}\right)^{\frac{2a}{a+1}} \sum_{i=1}^n |\varepsilon_i|^{2a} \quad \mbox{(by Lemma \ref{lemma_wanglong_lemma})} \\ \nonumber
    &=\left(\frac{1}{a+1}\right)^{\frac{2a}{a+1}} \sum_{i=1}^n \sigma_i^2 \quad \mbox{(by $|\varepsilon_i|^{2a}=\sigma_i^2$)}\\
    &=\left(\frac{1}{a+1}\right)^{\frac{2a}{a+1}} \sigma^T\sigma.
\end{align}
From \eqref{eq_lambda1_DD_bound} and \eqref{eq_V_2aa1_bound} we have
\begin{align}\label{eq_item2_bound}
    \frac{\sigma^TD^TD\sigma}{V ^{\frac{2a}{a+1}}} \nonumber
    &\ge \frac{\underline{\lambda}_1(D^TD)\sigma^T\sigma}{\left(\frac{1}{a+1}\right)^{\frac{2a}{a+1}} \sigma^T\sigma} \\
    &= (a+1)^{\frac{2a}{a+1}} \underline{\lambda}_1(D^TD)
\end{align}
for all $\varepsilon\in\Omega(c)\setminus\{0\}$.
Because $\lambda_1(D^TD)=\lambda_1(D^2)=\min_{i}((g_i^{\perp})^Tg_{i-1})^2\le1$, we have $\underline{\lambda}_1(D^TD)\le1$.

Second, by the definition of incidence matrices, $E$ is an incidence matrix of a directed and connected circular graph.
By \cite[Theorem 8.3.1]{graphbook}, we have $\mathrm{rank}(E)=n-1$ and consequently $\mathrm{rank}(E^TE)=n-1$.
Note $E\one=0$. Then $\Null(E^TE)=\mathrm{span}\{\one\}$.

Denote $\delta_i=\theta_i-\theta_i^*$ and
\begin{align}
    w_i=\frac{\cos \theta_i - \cos \theta_i^*}{\theta_i-\theta_i^*}.
\end{align}
Hence $\varepsilon_i=w_i\delta_i$.
Since $\lim_{\theta_i\rightarrow\theta_i^*} w_i=-\sin\theta_i^*$, the equation $\varepsilon_i=w_i\delta_i$ is still valid even if $\theta_i-\theta_i^*=0$.
Then we have
\begin{align}
    \varepsilon=W\delta,
\end{align}
where $W=\mathrm{diag}\{w_1,\dots,w_n\}\in\mathbb{R}^{n\times n}$ and $\delta=[\delta_1,\dots,\delta_n]^T\in\mathbb{R}^n$.
There exists sufficiently small $c$ such that $\theta_i(0)$ is sufficiently close to $\theta_i^*$ and hence $\theta_i,\theta_i^*\in(0,\pi)$ or $\theta_i,\theta_i^*\in(\pi,2\pi)$ for all $\varepsilon\in\Omega(c)$.
Clearly $w_i<0$ when $\theta_i,\theta_i^*\in(0,\pi)$, and $w_i>0$ when $\theta_i,\theta_i^*\in(\pi,2\pi)$.
Recall $(g_i^{\perp})^Tg_{i-1}>0$ when $\theta_i\in(0,\pi)$, and $(g_i^{\perp})^Tg_{i-1}<0$ when $\theta_i\in(\pi,2\pi)$ as shown in Lemma \ref{lemma_gi_perp} (iii).
Thus $(g_i^{\perp})^Tg_{i-1}w_i<0$ for all $i\in\{1,\dots,n\}$ and consequently the diagonal entries of $DW$ are with the same sign.
Moreover, because $\sum_i^n \theta_i\equiv\sum_i^n \theta_i^*$, the nonzero entries in $\delta$ are not with the same sign.
Therefore, the nonzero entries of $D\varepsilon=DW\delta$ are not with the same sign.
Furthermore, because $\sigma_i$ have the same sign as $\varepsilon_i$, the nonzero entries of $D\sigma$ are not with the same sign either.
By Lemma \ref{lemma_infimum_angle}, we have $D\sigma/\|D\sigma\|\in\mathcal{U}$ and
\begin{align}\label{eq_item1_bound}
    \frac{\sigma^TD^TE^TED\sigma}{\sigma^TD^TD\sigma}
    =\left(\frac{D\sigma}{\|D\sigma\|}\right)^T E^TE \left(\frac{D\sigma}{\|D\sigma\|}\right)
    >\frac{\lambda_2(E^TE)}{n}.
\end{align}
Because $\sum_{i=2}^n\lambda_i(E^TE)=\tr(E^TE)=\sum_{i,j=1}^n [E]_{ij}^2=2n$ and $\lambda_2(E^TE)$ is the smallest positive eigenvalue of $E^TE$, we have $(n-1)\lambda_2(E^TE)\le 2n$.
Thus $\lambda_2(E^TE)/n\le2/(n-1)\le1$ as $n\ge3$ and the equality holds only if $n=3$.

Substituting \eqref{eq_item2_bound} and \eqref{eq_item1_bound} into \eqref{eq_threeiterms} and \eqref{eq_V_dot_4} yields
\begin{align}\label{eq_V_dot_inequality}
    \dot{V}  \nonumber
    &\le -\frac{1}{\sum_{i=1}^n \|e_i\|}\sigma^TD^TE^TED\sigma \\
    &\le -\frac{1}{\sum_{i=1}^n \|e_i\|}K V ^{\frac{2a}{a+1}},
\end{align}
where
\begin{align}
    K = (a+1)^{\frac{2a}{a+1}} \underline{\lambda}_1(D^TD)\frac{\lambda_2(E^TE)}{n}.
\end{align}
As $a\in(0,1]$, $(a+1)^{\frac{2a}{a+1}}\le2$.
Recall $\underline{\lambda}_1(D^TD)\le1$ where the equality holds only if $(g_i^\perp)^Tg_{i-1}=\pm1$, i.e., $g_i\perp g_{i-1}$ for all $i\in\{1,\dots,n\}$; and $\lambda_2(E^TE)/n\le1$ where the equality holds only if $n=3$.
Obviously $g_i\perp g_{i-1}$ for all $i$ will not hold for a triangle with $n=3$.
Hence $\lambda_2(E^TE)/n=1$ and $\underline{\lambda}_1(D^TD)=1$ cannot hold simultaneously.
Then $\underline{\lambda}_1(D^TD)\lambda_2(E^TE)/n<1$ and hence $K \in(0,2)$.

\emph{Step 2:}

Since the distances between agents are not controlled directly, we cannot simply rule out the possibility that $\sum_{i=1}^n \|e_i\|$ may go to infinity.
Next we will prove $\sum_{i=1}^n\|e_i\|$ is bounded from upper by a finite positive value and hence ${1}/{\sum_{i=1}^n\|e_i\|}$ is bounded from below by a positive constant.

Recall $\dot{e}_i=\sigma_{i+1}(g_{i+1}-g_i)+\sigma_i(g_{i-1}-g_i)$ as shown in \eqref{eq_e_dot}. Denote $\rho=\sum_{i=1}^n\|e_i\|$. Then the time derivative of $\rho$ is
\begin{align}
    \dot{\rho}
\nonumber &=\sum_{i=1}^n \frac{\mathrm{d}\|e_i\|}{\mathrm{d}t} \\
\nonumber &=\sum_{i=1}^n g_i^T\dot{e}_i \\ \nonumber
\nonumber &=\sum_{i=1}^n g_i^T[\sigma_{i+1}(g_{i+1}-g_i)+\sigma_i(g_{i-1}-g_i)] \\ \nonumber
\nonumber &=\sum_{i=1}^n [\sigma_{i+1}(g_i^Tg_{i+1}-1)+\sigma_i(g_i^Tg_{i-1}-1)] \\
\nonumber &=\sum_{i=1}^n \sigma_{i+1}(g_i^Tg_{i+1}-1)+\sum_{i=1}^n \sigma_i(g_i^Tg_{i-1}-1) \\
\nonumber &=\sum_{i=1}^n \sigma_{i}(g_{i-1}^Tg_{i}-1)+\sum_{i=1}^n \sigma_i(g_i^Tg_{i-1}-1) \\
\nonumber &=2\sum_{i=1}^n \sigma_i(g_i^Tg_{i-1}-1) \\
    &=2v^T\sigma,
\end{align}
where $v=[v_1,\dots,v_n]^T\in\mathbb{R}^n$ with $v_i=g_i^Tg_{i-1}-1$. Then by the Cauchy-Schwarz inequality
\begin{align}\label{eq_rho_dot_abs}
    |\dot{\rho}| =2|v^T\sigma|
    \le 2\|v\|\|\sigma\|
    \le 2\beta\|\sigma\|,
\end{align}
where $\beta$ is the maximum of $\|v\|$ over the compact set $\Omega(c)$.

For any $c\le1$, we have $V\le1$ and then $V ^{\frac{2a}{1+a}}\ge V $ as ${2a}/{(1+a)}\le1$ for all $\varepsilon\in\Omega(c)$.
Thus
\begin{align}
    \dot{V}
    \le -\frac{K }{\rho}V ^{\frac{2a}{1+a}}
    \le -\frac{K }{\rho}V .
\end{align}
By the comparison lemma \cite[Lemma 3.4]{Khalilbook}, the above equation implies
\begin{align}
    V(t)\le V(0)+\int_0^t -\frac{K}{\rho(\tau)}V(\tau) \mathrm{d}\tau.
\end{align}
Applying the Gronwall-Bellman inequality \cite[Lemma A.1]{Khalilbook} to the above equation gives
\begin{align}\label{eq_V_inequality}
    V (t)\le V (0)\exp\left(\int_0^t -\frac{K }{\rho(\tau)} \mathrm{d}\tau\right).
\end{align}
In addition, we have
\begin{align}
    V ^{\frac{2a}{1+a}} \nonumber
    &=\left(\frac{1}{a+1}\right)^{\frac{2a}{a+1}} \left(\sum_{i=1}^n |\varepsilon_i|^{a+1}\right)^{\frac{2a}{a+1}} \\ \nonumber
    &\ge \left(\frac{1}{a+1}\right)^{\frac{2a}{a+1}} \frac{1}{n^{\frac{1-a}{1+a}}} \sum_{i=1}^n |\varepsilon_i|^{2a} \quad \mbox{(By Lemma \ref{lemma_wanglong_lemma})}\\
    &=\left(\frac{1}{a+1}\right)^{\frac{2a}{a+1}} \frac{1}{n^{\frac{1-a}{1+a}}} \|\sigma\|^2,
\end{align}
which implies
\begin{align} \label{eq_sigma_norm_bound}
    \|\sigma\|^2
    &\le \underbrace{(a+1)^{\frac{2a}{a+1}}n^{\frac{1-a}{1+a}}}_{\kappa} V ^{\frac{2a}{a+1}}.
\end{align}

Substituting \eqref{eq_sigma_norm_bound} and \eqref{eq_V_inequality} into \eqref{eq_rho_dot_abs} yields
\begin{align}
    |\dot{\rho}| \nonumber
    &\le 2\beta\|\sigma\| \\ \nonumber
    &\le 2\beta\sqrt{\kappa}V ^{\frac{a}{a+1}} \\ \nonumber
    &\le 2\beta\sqrt{\kappa}V (0)^{\frac{a}{a+1}}\exp\left(\int_0^t -\frac{{\frac{a}{a+1}}K }{\rho(\tau)} \mathrm{d}\tau\right) \\
    &\le \exp\left(\int_0^t -\frac{{\frac{a}{a+1}}K }{\rho(\tau)} \mathrm{d}\tau\right).
\end{align}
The last inequality uses the fact that there exists a sufficiently small $c$ such that $2\beta\sqrt{\kappa}V (0)^{\frac{a}{a+1}}\le1$ when $V (0)\le c$.
Because $a/(1+a)\in(0,0.5]$ and $K \in(0,2)$, we have ${a}K /{(a+1)}\in(0,1)$.
Then by Lemma \ref{lemma_gammaupperbound}, $\rho$ is bounded from upper by a finite constant. Denote this upper bound as $\gamma $.
Then $\dot{V}$ in \eqref{eq_V_dot_inequality} becomes
\begin{align}\label{eq_V_dot_final}
    \dot{V}  \le -\frac{K }{\gamma }V ^{\frac{2a}{1+a}}.
\end{align}

For ease of presentation, we make a temporary assumption here that no collision between any agents occurs for all $t\in[0,+\infty)$.
Without this assumption, inequality \eqref{eq_V_dot_final} only holds until collision happens.
In the next section we will prove that the proposed control law ensures collision avoidance.
Then this assumption can be removed.
Under the collision avoidance assumption, inequality \eqref{eq_V_dot_final} holds for all $t\in[0,+\infty)$.

If $a\in(0,1)$, $2a/(1+a)\in(0,1)$. By \cite[Theorem 4.2]{bernstein2000}, the solution to \eqref{eq_controlLaw} starting from $\Omega(c)$ converges to the equilibrium $\varepsilon=0$ in finite time.

If $a=1$, $2a/(1+a)=1$. By \cite[Theorem 3.1]{lyapunovbook}, the equilibrium $\varepsilon=0$ is locally exponentially stable.
\end{proof}

\section{Formation Behaviors in the Plane}\label{section_formationbehavior}
Because the target formation is only constrained by angles, the positions of the agents and the inter-agent distances are not controlled in the final converged formation.
In addition to the dynamics of $\varepsilon$ as analyzed in the last section, it is also important to investigate the behavior of the positions of the agents in the plane.

From control law \eqref{eq_controlLaw}, one trivial behavior of the formation is that $\dot{z}=0$ if $\varepsilon=0$.
In other words, we can rule out the possibility that the agents are still moving while the target angles have been achieved.
Another trivial behavior of the formation is that $\dot{z}=0$ if $\dot{\varepsilon}=0$.
That is because $\dot{V}=0$ if and only if $\varepsilon=0$.
Intuitively speaking, it is impossible that control law \eqref{eq_controlLaw} only changes the orientation, translation or scale of the formation while keeping all angles unchanged.

Collision avoidance is an important issue in all kinds of formation control.
It is especially important for formation control using bearing-only measurements because the distance between any two agents cannot be measured or controlled directly.
The next result shows a simple but important behavior of the formation by the proposed control law: when the initial angles are sufficiently close to the target angles, the final converged positions of the agents are also sufficiently close to the initial positions.
As a consequence, the proposed control law can locally ensure collision avoidance.
Then the temporary assumption on collision avoidance in the proof of Theorem \ref{theorem_stability} can be removed.

\begin{my_theorem}\label{theorem_closetoinitial}
    Suppose in the initial formation $z_j(0)\ne z_k(0)$ for all $j,k\in\{1,\dots,n\}$ and $j\ne k$.
    Then there exists a sufficiently small $\|\varepsilon(0)\|_{a+1}$ and a positive constant $\eta$ such that the distance between $z(t)$ and $z(0)$ as shown below satisfies
    \begin{align}
        \sum_{i=1}^n \|z_i(t)-z_i(0)\|
        \le \eta\|\varepsilon(0)\|_{a+1}
    \end{align}
    for all $t\in[0,+\infty)$.
    As a result, there exists a sufficiently small $\|\varepsilon(0)\|_{a+1}$ such that collision avoidance between any agents can be ensured by control law \eqref{eq_controlLaw}.
\end{my_theorem}
\begin{proof}
    We prove by contradiction.
    Suppose no collision between any agents occurs until $T^*\in(0,+\infty)$.
    As shown in the proof of Theorem \ref{theorem_stability}, there exists a sufficiently small $c>0$ such that inequality \eqref{eq_V_dot_final} holds for $t\in[0,T]$ with $T<T^*$ if $\varepsilon(0)\in\Omega(c)$.
    When $c$ is sufficiently small, we also have $V(0)\le1$ and hence $V(t)\le1$ for all $t\in[0,T]$.
    Since $2a/(1+a)\in(0,1]$, we have $V ^{\frac{2a}{1+a}}\ge V$ and consequently
    \begin{align}
        \dot{V}  \le -\frac{K }{\gamma }V ^{\frac{2a}{1+a}} \le -\frac{K }{\gamma }V,
    \end{align}
    which implies
    \begin{align} \label{eq_V_upperbound}
        V(t) \le V(0)e^{-\frac{K }{\gamma}t}
    \end{align}
    for all $t\in[0,T]$.
    Substituting $V=1/(a+1)\|\varepsilon\|_{a+1}^{a+1}$ into \eqref{eq_V_upperbound} yields
    \begin{align}\label{eq_xi_a1norm}
        \|\varepsilon(t)\|_{a+1} \le \|\varepsilon(0)\|_{a+1} e^{-\frac{K}{(a+1)\gamma}t}.
    \end{align}
    Thus the following ``distance'' between $z(t)$ and $z(0)$ satisfies
    \begin{align}\label{eq_initialandfinaldistance}
        \sum_{i=1}^n \|z_i(t)-z_i(0)\| \nonumber
        &=\sum_{i=1}^n \left\|\int_0^t \varepsilon_i(g_i-g_{i-1})\mathrm{d}\tau\right\| \quad \mbox{(by control law \eqref{eq_controlLaw})} \\ \nonumber
        &\le \sum_{i=1}^n \int_0^t |\varepsilon_i|\|g_i-g_{i-1}\|\mathrm{d}\tau \\ \nonumber
        &\le 2 \int_0^t \sum_{i=1}^n |\varepsilon_i|\mathrm{d}\tau \quad \mbox{(becuause $\|g_i-g_{i-1}\|\le\|g_i\|+\|g_{i-1}\|=2$)} \\ \nonumber
        &= 2 \int_0^t \|\varepsilon\|_1\mathrm{d}\tau \\ \nonumber
        &\le 2C \int_0^t \|\varepsilon\|_{a+1}\mathrm{d}\tau \quad \mbox{(by Lemma \ref{lemma_vector_norm_equivalent})}\\ \nonumber
        &\le 2C \int_0^t \|\varepsilon(0)\|_{a+1} e^{-\frac{K}{(a+1)\gamma}\tau} \mathrm{d}\tau \quad \mbox{(by \eqref{eq_xi_a1norm})}\\ \nonumber
        &=2C\|\varepsilon(0)\|_{a+1}\frac{(a+1)\gamma}{K}\left(1-e^{-\frac{K}{(a+1)\gamma}t}\right) \\
        &\le \underbrace{\frac{2C(a+1)\gamma}{K}}_{\eta}\|\varepsilon(0)\|_{a+1}
    \end{align}
    for all $t\in[0,T]$.

    Now suppose agents $j$ and $k$ collide at time $T^*$. Then $z_j(T^*)=z_k(T^*)$ and
    \begin{align}\label{eq_conflict1}
        \sum_{i=1}^n \|z_i(T^*)-z_i(0)\| \nonumber
        & \ge \|z_j(T^*)-z_j(0)\| + \|z_k(T^*)-z_k(0)\| \\ \nonumber
        & \ge \|z_j(T^*)-z_j(0) - z_k(T^*)+z_k(0)\|\\
        & = \|z_k(0)-z_j(0)\|.
    \end{align}
    However, given any positive constant $r$ with $r< \|z_k(0)-z_j(0)\|$, by \eqref{eq_initialandfinaldistance} we can always find a sufficiently small $\varepsilon(0)$ such that
    \begin{align}\label{eq_conflict2}
        \sum_{i=1}^n \|z_i(t)-z_i(0)\| \nonumber
        &\le \eta \|\varepsilon(0)\|_{a+1} \\
        &< \|z_k(0)-z_j(0)\|-r
    \end{align}
    for all $t\in[0,T]$ and all $T< T^*$.
    Clearly \eqref{eq_conflict1} conflicts with \eqref{eq_conflict2}.
    Therefore, collision avoidance can be guaranteed by control law \eqref{eq_controlLaw} if $\varepsilon(0)$ is sufficiently small.
    Consequently \eqref{eq_V_dot_final} and \eqref{eq_initialandfinaldistance} holds for all $t\in[0,+\infty)$.
\end{proof}

\section{Simulations}\label{section_simulation}
In this section, we present numerical simulations to illustrate our theoretical analysis.

\begin{figure}
  \centering
  \subfloat[Formation evolution with $a=1$.]{\includegraphics[width=0.5\linewidth]{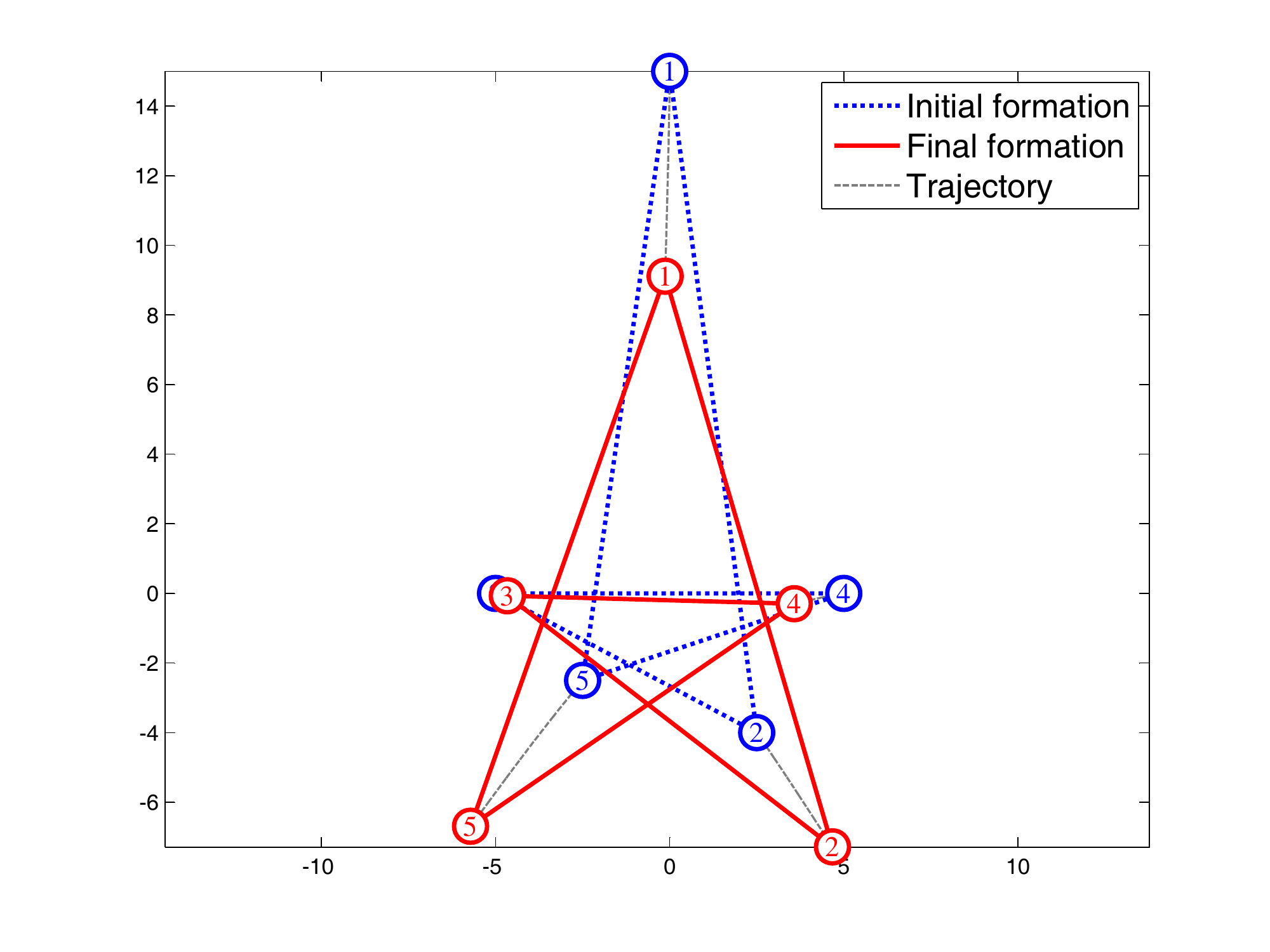}}
  \subfloat[Angle error and Lyapunov function evolution with $a=1$.]{\includegraphics[width=0.5\linewidth]{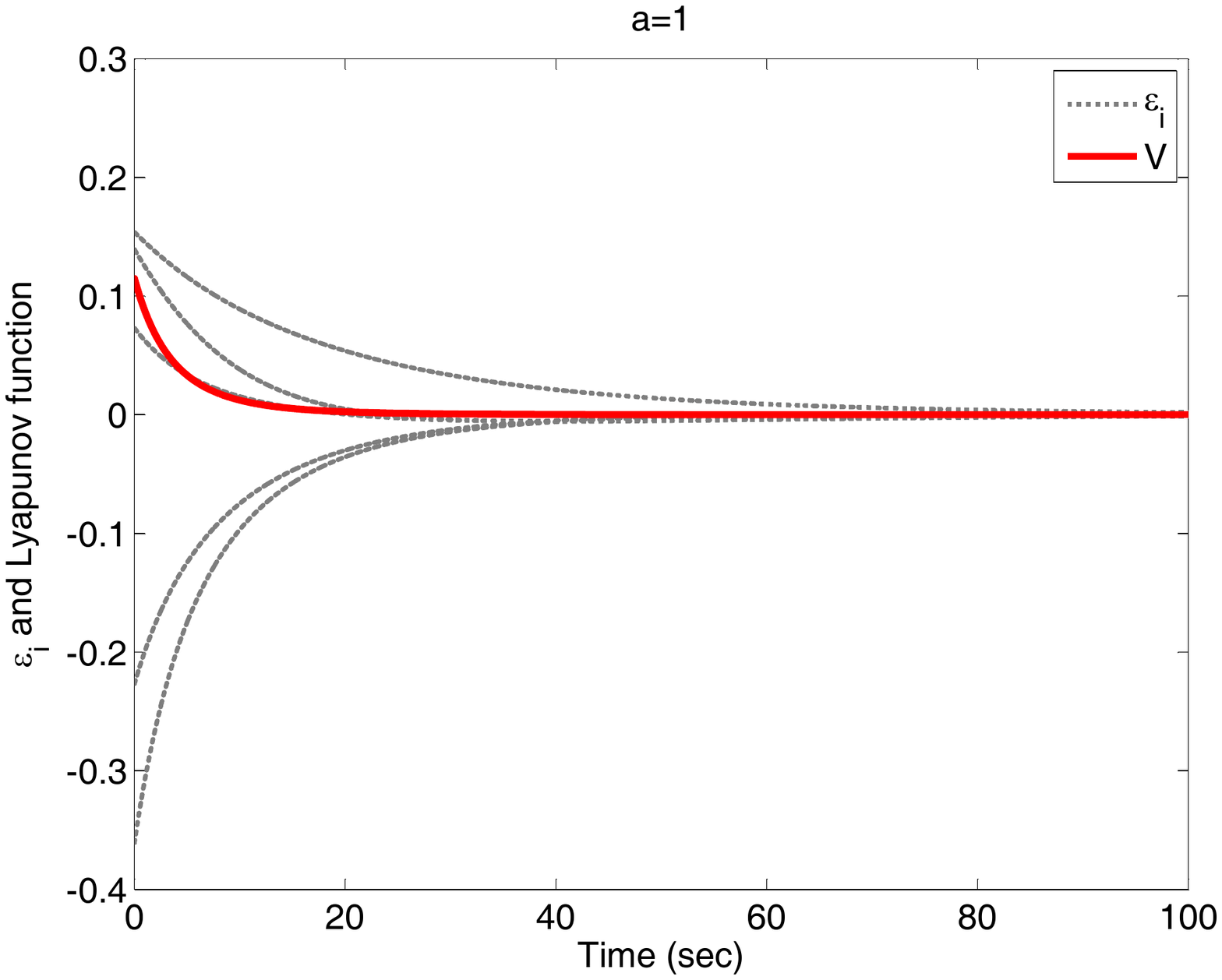}} \\
  \subfloat[Formation evolution with $a=0.3$.]{\includegraphics[width=0.5\linewidth]{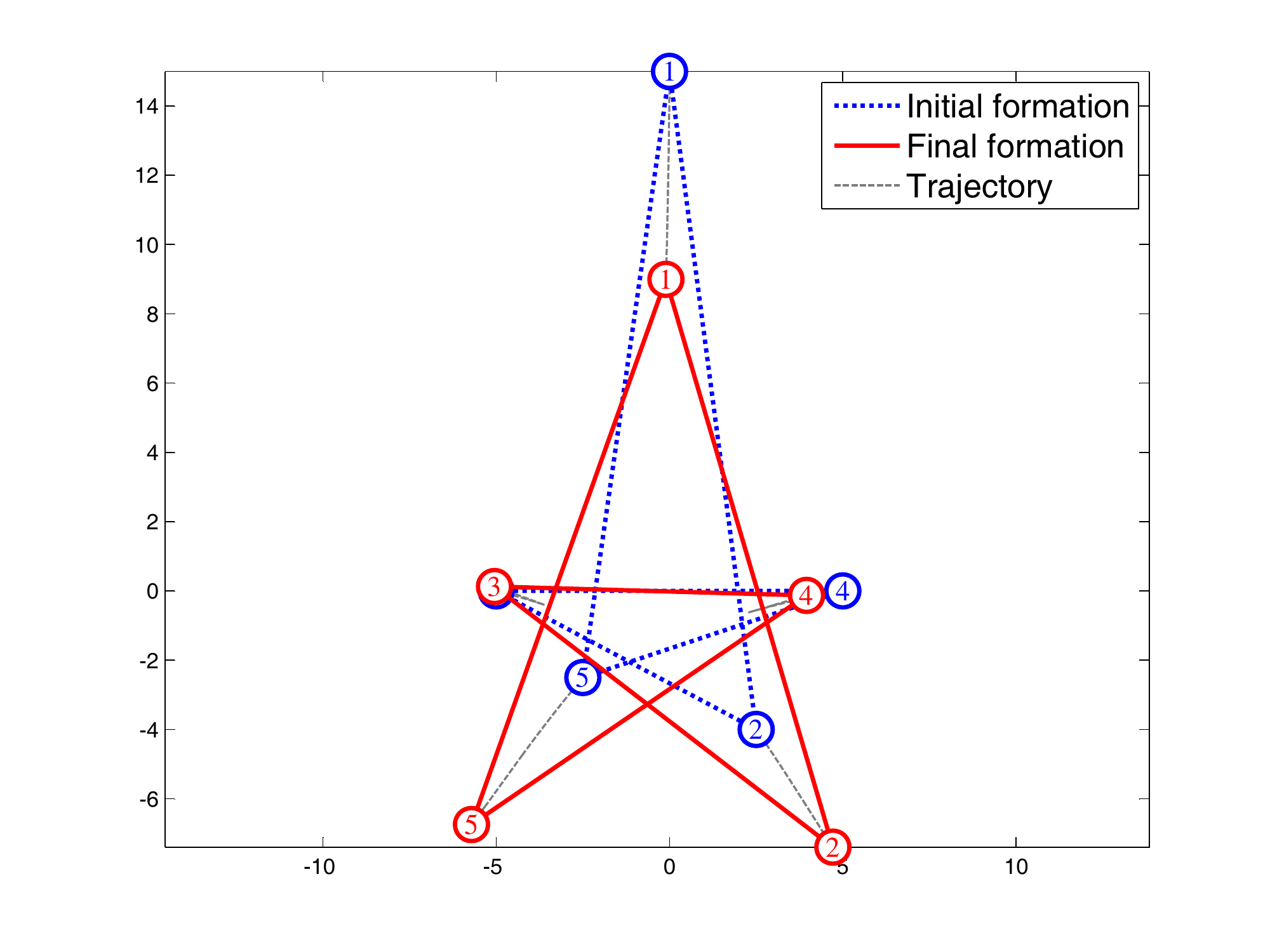}}
  \subfloat[Angle error and Lyapunov function evolution with $a=0.3$.]{\includegraphics[width=0.5\linewidth]{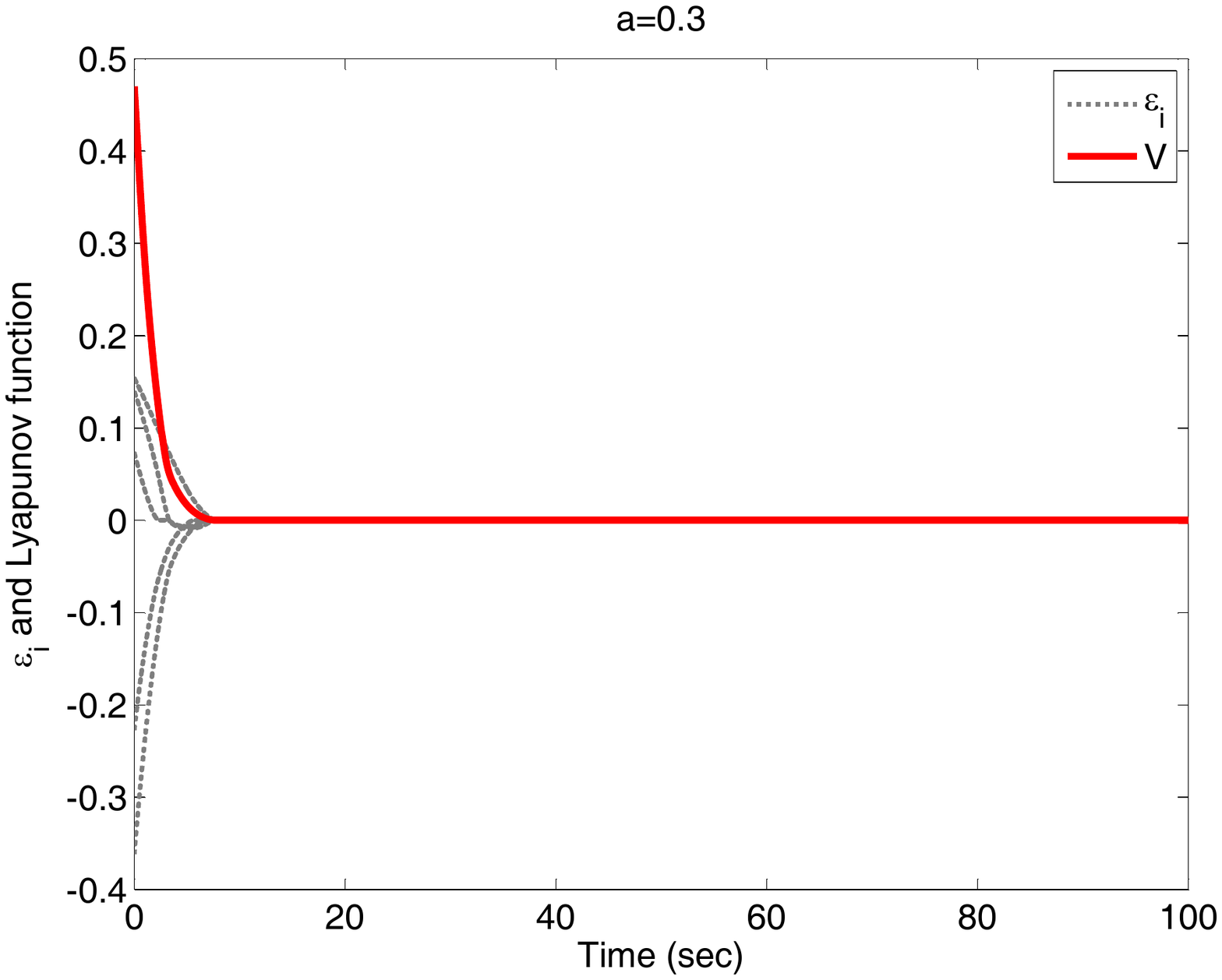}}
  \caption{Control results by the proposed control law with $n=5$ and $\theta_1^*=\dots=\theta_n^*=36$ deg.}
  \label{fig_simulation_five}
\end{figure}

The target formation in Fig. \ref{fig_simulation_five} is a non-convex star polygon with $n=5$.
The angle at each vertex in the target formation is $\theta_i^*=36$ deg for all $i\in\{1,\dots,5\}$.
From an initial formation, the proposed control law effectively reduces the angle errors and steers the formation to a target formation.
Because the target formation is constrained only by angles, the shape of the final formation does not look like a regular star polygon though all angle errors achieve zero.

\begin{figure}
  \centering
  \subfloat[Formation evolution with $a=1$.]{\includegraphics[width=0.5\linewidth]{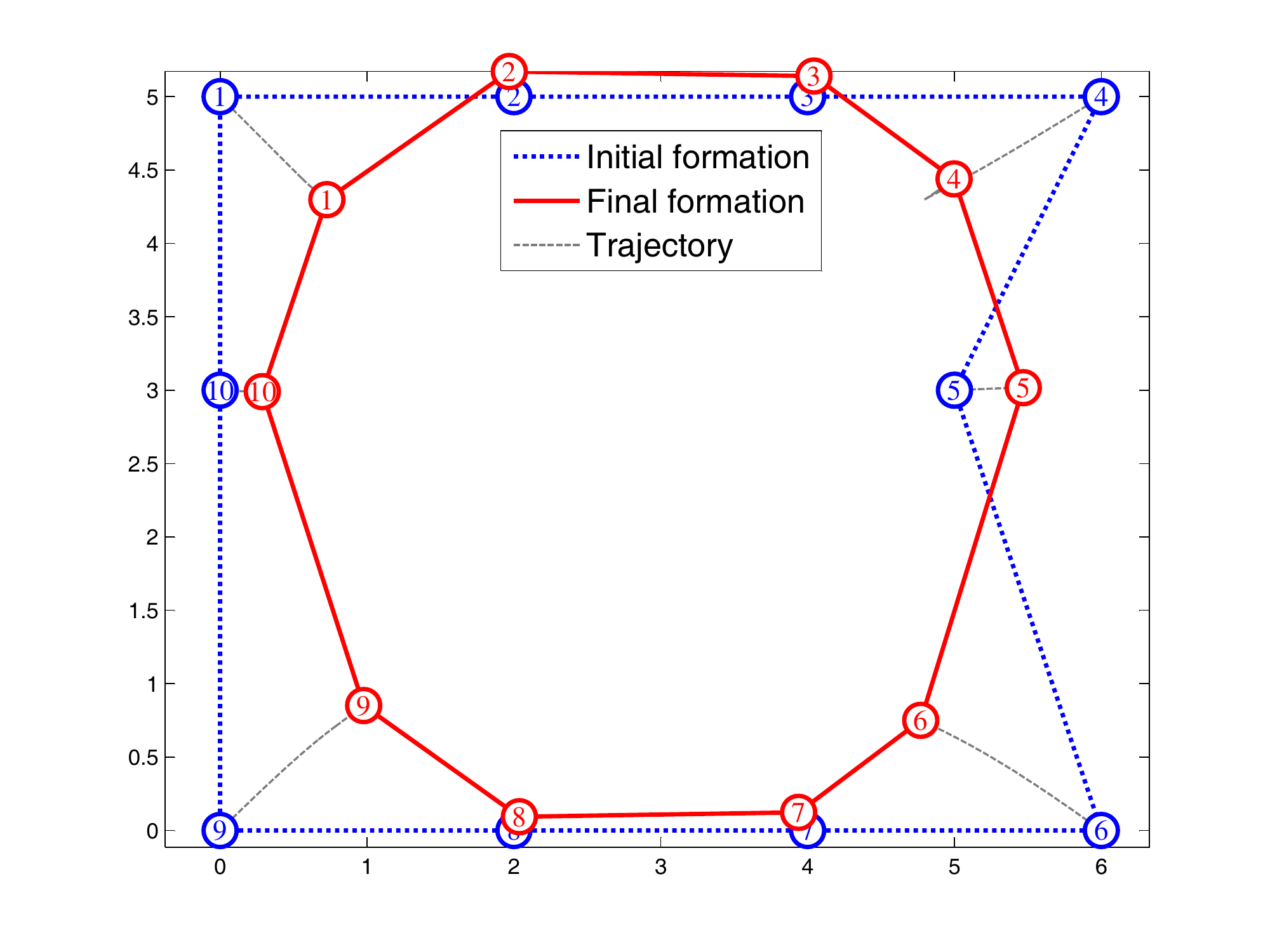}}
  \subfloat[Angle error and Lyapunov function evolution with $a=1$.]{\includegraphics[width=0.5\linewidth]{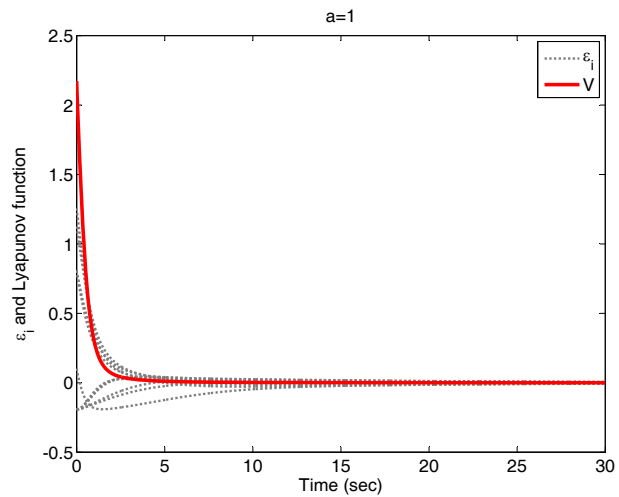}} \\
  \subfloat[Formation evolution with $a=0.6$.]{\includegraphics[width=0.5\linewidth]{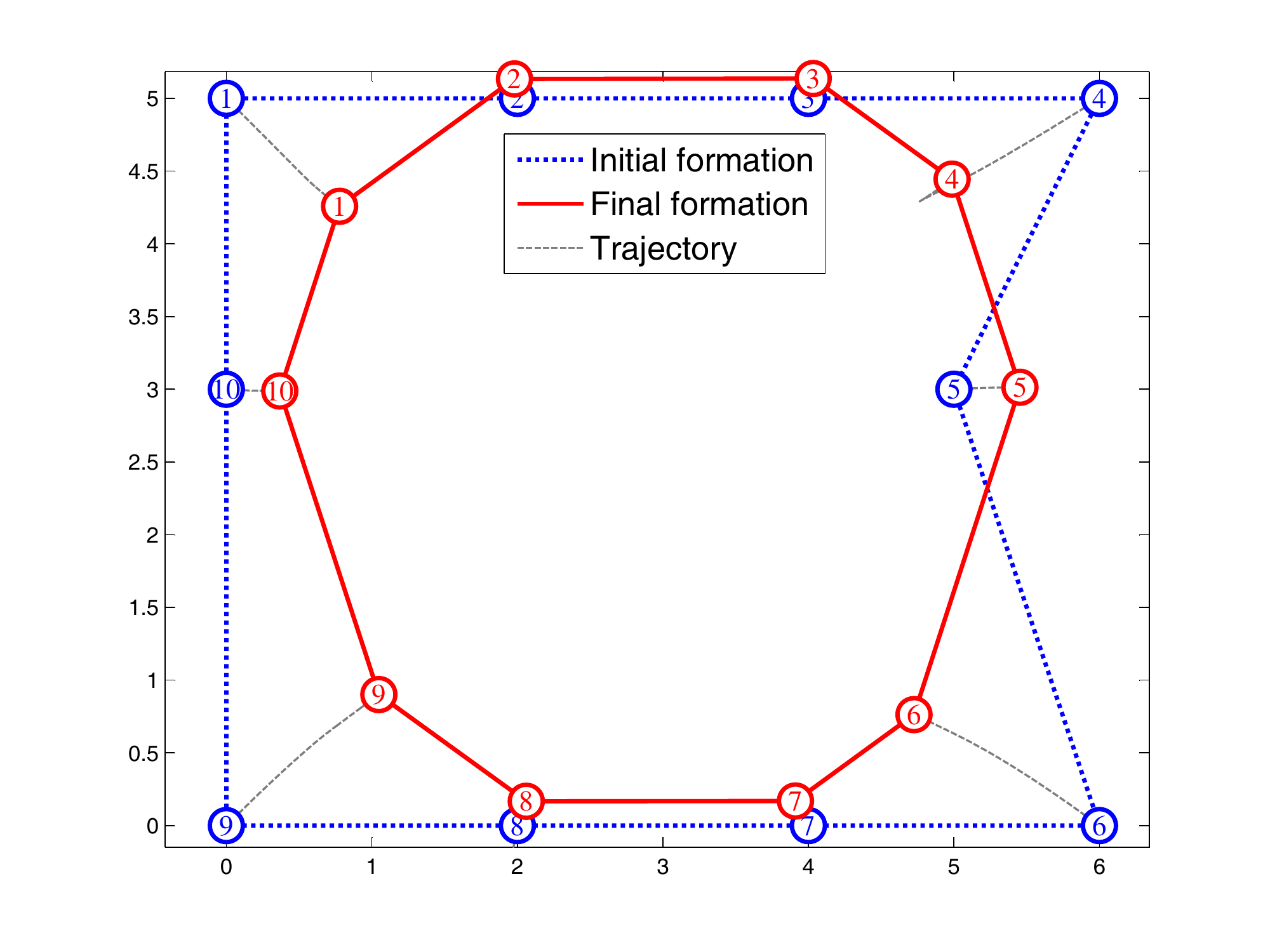}}
  \subfloat[Angle error and Lyapunov function evolution with $a=0.6$.]{\includegraphics[width=0.5\linewidth]{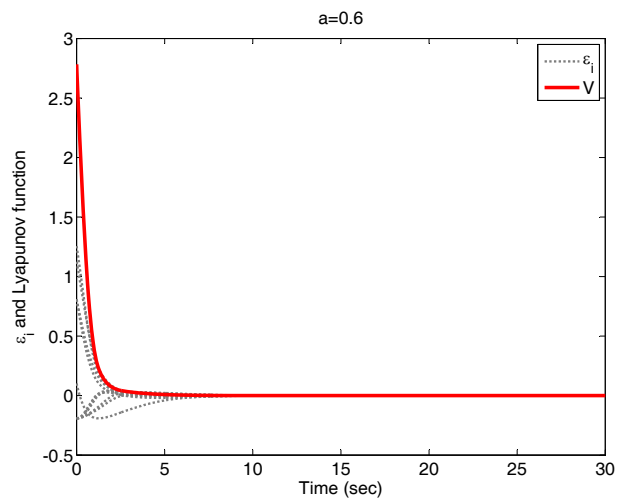}}
  \caption{Control results by the proposed control law with $n=10$ and $\theta_1^*=\dots=\theta_n^*=144$ deg.}
  \label{fig_simulation_ten}
\end{figure}

The target formation in Fig. \ref{fig_simulation_ten} is a ten-side polygon, where the angle at each vertex is $\theta_i^*=(10-2)\times180/10=144$ deg for all $i\in\{1,\dots,10\}$.
In the stability analysis, we have mentioned there exists a sufficiently small $c$ such that $\theta_i,\theta_i^*\in(0,\pi)$ or $(\pi,2\pi)$ for all points in $\Omega(c)$.
However, it is notable that in the initial formation $\theta_i(0)=\pi$ for $i=2,3,7,8,10$ and $\theta_5(0)\in(\pi,2\pi)$ but $\theta_5^*\in(0,\pi)$. The target formation can still be achieved.
Hence from the simulation, the attractive region of the target formation by the proposed control law is not necessarily small.

Figures \ref{fig_simulation_five} and \ref{fig_simulation_ten} show that the angle errors and the Lyapunov function converge to zero in finite time if $a<1$.

\section{Conclusion}\label{section_conclusion}
This paper addresses a relatively new formation control problem: distributed control of angle-constrained circular formations using local bearing measurements.
The proposed control law ensures exponential or finite-time convergence.
The presented Lyapunov analysis might be helpful for future research on more complicated target formations.

The work in this paper is the first step towards the control of generic angle-constrained formations using bearing-only measurements.
Many interesting problems in this field are still unsolved.
In the future, one immediate research plan is to study the control of generic angle-constrained target formations, in which one agent may correspond to multiple constrained angles.
It is also meaningful to study formation control involving agent position estimation based on bearing measurements.

\bibliography{zsybib}
\bibliographystyle{ieeetr}

\end{document}